\documentclass[twocolumn,showpacs,preprintnumbers,amsmath,amssymb]{revtex4}

\usepackage{graphicx}
\usepackage{dcolumn}
\usepackage{bm}

\begin{document}

\title{Charge Inversion of Divalent Ionic Solutions in Silica Channels}

\author{Christian D.\ Lorenz and Alex Travesset}
\affiliation{Department of Physics and Astronomy and Ames
Laboratory, Iowa State University, Ames, Iowa 50011}

\date{\today}

\begin{abstract}

Recent experiments (F.H.J. Van Der Heyden et al., PRL {\bf 96},
224502 (2006)) of streaming currents in silica nanochannels with
divalent ions report charge inversion, i.e. interfacial charges
attracting counterions in excess of their own nominal charge, in
conflict with existing theoretical and simulation results. We
reveal the mechanism of charge inversion by using all-atomic
molecular dynamics simulations. Our results show excellent
agreement with experiments, both qualitatively and quantitatively.
We further discuss the implications of our study for the general
problem of ionic correlations in solutions as well as in regards
of the properties of silica-water interfaces.

\end{abstract}

 \maketitle

\section{Introduction}

Counterions in aqueous solution play a crucial role in the
self-assembly of colloids and polymers, cell signaling,
microfluidics and fuel cells. In recent years, there has been a
considerable body of experimental and theoretical work aimed at
understanding the rich and complex phenomenology of ions in
aqueous solution and their interactions with charged interfaces
\cite{McLaughlin1989,Grosberg2002,Levin2002,Boroud2005}. A
representative example of this phenomenology is charge inversion
(CI), where interfacial charges attract counterions in excess of
their own nominal charge, thus leading to an interface whose
effective surface charge is reversed.  The first simulation study that showed evidence of CI with divalent ions was conducted by Torrie and Valleau \cite{torriejpc82},  in which the authors used a primitive hard sphere model for the ions and a continuum solvent model with a dielectric constant.  Still the origins of CI are poorly understood, as highlighted in a recent review article by
Lyklema \cite{Lyklema2006}. In particular, it is unclear to what extent existing
theories explain the observed CI reported in experimental work.

\begin{figure}[ctb]
\begin{center}
\includegraphics[width=3in]{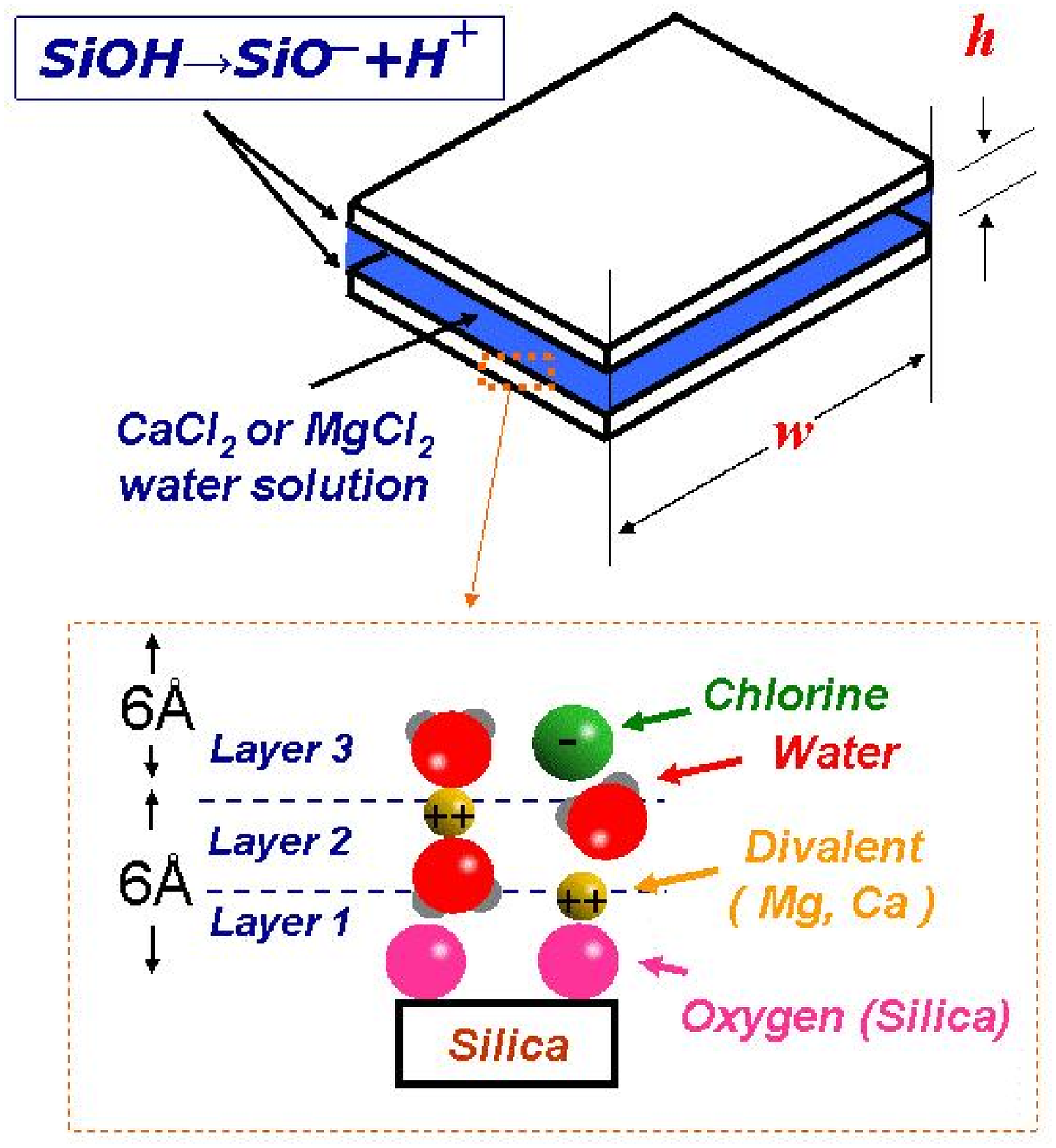}
\caption{(Color online:) Cartoon picture of a silica nanochannel.
Very near to the silica interface, we identify a first layer of
bound counterions (layer 1), a layer of hydrated counterions
(layer 2) and a third layer (layer 3), which corresponds to the
diffuse layer and exhibits CI. }\label{fig:cartoon}.
\end{center}
\end{figure}

A recent experiment has provided a detailed investigation of CI by
multivalent counterions from an analysis of streaming currents,
the electric currents resulting from applying a pressure gradient
along a charged nanofluidic channel containing electrolyte
solutions (see Fig.~\ref{fig:cartoon}) \cite{Lemay2006}. The
precision and accuracy of such experiments provide a perfect
arena to test existing theoretical models of CI in
particular and the more general problem of ion distributions.

Different theories for CI have been proposed over the years. In
Ref.~\cite{Shklovskii1999} it was proposed that CI arises from the
gain in correlation energy brought about by lateral correlations
(LC) among mobile counterions. For divalent ions in silica
nanochannels, this theory would predict CI at $c_{inv}\sim 10$ mM
\footnote{The plasma parameter $\Gamma$ may be too small for this
theory to rigorously apply \cite{Grosberg2002}.}-an order of
magnitude lower than what is observed experimentally
\cite{Lemay2006}. In Refs.\ \cite{Travesset2006,Faraudo2006} it has been
pointed out that the intrinsic discreteness of interfacial charges
may give rise to CI by transverse correlations (TC), that is,
correlations between counterions and interfacial charges. In
particular, it has been shown that binding constants of divalent
ions to oxygen atoms are typically of the order of $K_B\sim 10$~
M$^{-1}$, thus implying CI at $c_{inv}\sim 1/K_B\approx 0.1$~M.
Numerical simulations with primitive models, where the solvent is
considered implicitly as a medium of dielectric constant
$\varepsilon$, have been
reported\cite{Trulsson2006,Molina2006,Diehl2006}, and show general
agreement with LC theories, thus implying lower than the
$c_{inv}\approx 400$~mM observed in experiments \cite{Lemay2006}.
Other theoretical work has focused in the computation of
mobilities \cite{Lozada2001} as a function of $\zeta$-potential,
but as reviewed in \cite{Lyklema2006}, the interpretation of experimentally determined $\zeta$-potentials is not entirely clear. A clear understanding of
the CI phenomenon may therefore require approaches where water is
considered explicitly and the interface is modelled realistically,
with more structure than a uniform surface charge. As discussed in
\cite{Faraudo:2004,Faraudo2006}, these effects are expected to
play a critical role.

In this paper, we investigate the origin of CI with atomic
resolution by using molecular dynamics (MD) simulations of a charged
silica interface in contact with an aqueous solution with divalent
ions. This study is further motivated by recent interest in
understanding the structure of pure water in contact with silica
\cite{Engemann2004,Ostroverkhov2005,Aarts2005,Asay2005,Joseph2006},
as some evidence suggests the presence of highly ordered
interfacial water structures. It has not been investigated,
however, how divalent ions, which have a very cohesive hydration
sheath, may modify interfacial water. In addition, our results
have implications in the field of microfluidics because silica is
the most common material used in nano- and microfluidic
devices\cite{Stone2004}, including the recently synthesized
nanoporous functionalized silica thin films
\cite{Doshi2000,brinker} that are being investigated as
desalination membranes \cite{leungprl}.

\section{Simulation details}

We consider the system schematically shown in
Fig.~\ref{fig:cartoon} where an interface of amorphous silica is
in contact with an aqueous solution. Throughout this letter, we
will consider ionic strengths such that $\lambda_D<<h$, where
$\lambda_D$ is the Debye length and $h$ is the separation between
the two silica interfaces (see Fig.\ref{fig:cartoon}). We can
therefore neglect the interactions between the two silica interfaces
and perform our MD simulations on a single interface.

Amorphous silica was generated according to a previously reported
procedure \cite{chandross2004,lorenz2005}.  We start with a bulk
$\alpha$-quartz crystalline substrate that is heated to high
temperature ($\sim 2500$~K) and quenched to $300$~K. We then
generate two free surfaces in the z-dimension and further anneal
the substrate until the density of bond defects on each interface
is representative of the experimentally observed density on
amorphous silica ($4.0$-$5.0$/nm$^2$)\cite{zhuravlev}.  The silicon and oxygen
defects are terminated with -OH and -H respectively. The silica
interface is given a negative charge by randomly selecting -OH
groups and removing the corresponding proton, leaving a net
surface charge of $\sim 1e/110$~\AA$^{-2}$, in line with
experimental results \cite{Lemay2006}. The amorphous silica had a
total area of $68.7076 \times 68.082$ \AA$^2$ in contact with an
aqueous solution containing water ($\sim 8500$ water molecules)
and divalent ions. Water was modelled explicitly by using the
TIP3P model \cite{jorgensen1983}, the two cations used in the
simulation, (Ca$^{2+}$, Mg$^{2+}$), were modeled with Aqvist's
parameters \cite{aqvist1990} and Cl$^{-}$ with the OPLS forcefield
\cite{chandrasekhar1984}. The silica was modeled by the OPLS force
field described in \cite{jorgensen1996}.

The simulations reported in this article were performed with the
LAMMPS molecular dynamics code \cite{lammps}. In all simulations,
the temperature is fixed at T$\approx 300$ K and controlled with a
Nose-Hoover thermostat and a time constant of 0.01~fs$^{-1}$. The
SHAKE algorithm \cite{shake} was used to constrain the bond
lengths and angles of the water molecule and the bond lengths of
the OH bonds on the silica substrate surface.  A 1~fs time step
was used and the van der Waals (vdW) interaction is cut off at
10~\AA. A slab version of the particle-particle particle-mesh
algorithm \cite{crozier2001} is used to compute long-range
Coulombic interactions. An average run lasted around 12 ns, with
data being collected and analyzed over the last 8 ns. A typical
run took 4608 node-hours per 12 ns simulation.

\section{Simulation Results}
\begin{figure}[ctb]
\begin{center}
\includegraphics[width=3.4 in]{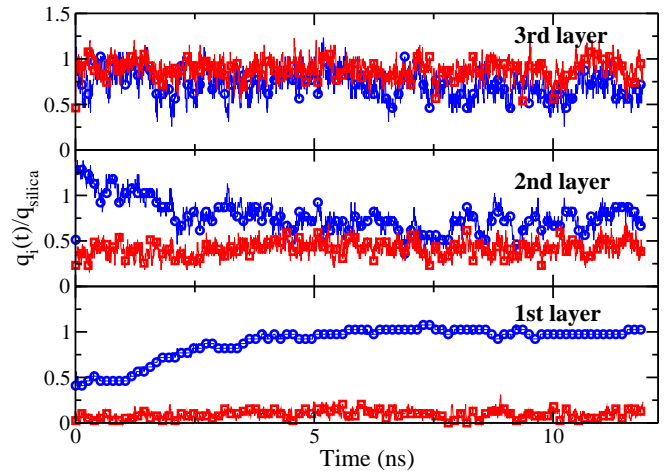}
\caption{(Color online:) Evolution of the Ca$^{2+}$ charge (blue
circle) and Cl$^{-}$ charge (red square) relative to the
interfacial charge as a function of time for the different layers.
Layer 1 represents bound oxygens (defined such that O-Ca$^{2+}$
distance is less than 3 \AA) layer 2 is defined by the size of the
hydrated divalent ions (O-Ca$^{2+}$ distance is within 3 and 6~
\AA) \ and layer 3 is the last layer before bulk values are
attained, with O-Ca$^{2+}$ distance $<12$~\AA. (see
Fig.~\ref{fig:cartoon} for a schematic definition of the different
layers). Results are for 1.0 M
concentrations.}\label{fig:time_series_T}.
\end{center}
\end{figure}

\begin{table}
\caption{Relative charge population of the ionic species of
Ca$^{2+}$ and Cl$^{-}$ in each layer.
Values in parentheses represent fluctuations in the last
digit(s).\label{table:rhocharge}}
\begin{tabular}{|c|ll|ll|ll|}
\hline
System &1$^{st}$ layer& &2$^{nd}$ layer&&3$^{rd}$ layer&\\
  & Ca$^{2+}$ & Cl$^-$ & Ca$^{2+}$ & Cl$^-$ & Ca$^{2+}$ & Cl$^-$ \\
\hline
0.5 M & $0.68(41)$ & $0.02(1)$ & $0.54(23)$ & $0.13(2)$ & $0.33(7)$ & $0.36(8)$ \\
1.0 M & $0.99(29)$ & $0.10(2)$ & $0.69(13)$ & $0.43(6)$ & $0.70(18)$ & $0.88(12)$ \\
1.5 M & $0.92(40)$ & $0.11(1)$ & $0.85(27)$ & $0.55(19)$ & $1.03(26)$ & $1.14(17)$ \\
\hline
\end{tabular}
\end{table}

It is convenient to divide the region near the silica interface
into three layers as depicted in Fig.~\ref{fig:cartoon}. The first
layer is defined so that it includes all partially dehydrated
divalent ions bound to silica oxygens. The second layer includes
all counterions within a distance of 6~\AA\ (measured from the
center of the silica oxygens) and basically consists of hydrated
counterions. The third layer extends up to a region of 12 \AA,
beyond which bulk density profiles are attained. We recall that in
textbooks in physical chemistry \cite{Lyklema1995}, layer 1 is
referred to as the Inner Helmholtz plane, layer 2 as the Outer
Helmholtz plane and layer three as the diffuse layer. The time
evolution for the population of both Ca$^{2+}$ and Cl$^{-}$ on the
different layers are shown in Fig.~\ref{fig:time_series_T} at 1.0 M
concentration. The population is expressed as a function of total
charge of the ionic species within the layer relative to the total
net interfacial charge. As is clear from
Fig.~\ref{fig:time_series_T}, the population of the different
layers reaches equilibrium after an initial period of about 3~ns.
We further measured residence times and found that the average
residence time for Ca$^{2+}$ ions were $\sim 2.0$~ns, $\sim
200$~ps and $\sim 30$~ps in the first, second and third layers
respectively. The total simulation time for our simulations was 12
ns thus providing ample time to compute equilibrium quantities. No
significant differences were observed for other CaCl$_2$
concentrations. A summary of the average populations within different layers is shown in Table~\ref{table:rhocharge}.

Results for MgCl$_2$, however, show significant
differences, as residence times in the first and second layer are
surprisingly long. As an example, within 12 ns, only a single
Mg$^{2+}$ ion was exchanged from the first to the second layer. It
is therefore possible that the first and second layer for MgCl$_2$
are not thermalized. The results for the MgCl$_2$ solutions are also of interest as will be clear, so in view of these caveats we will relegate a detailed summary of these results to Appendix  \ref{appendixa}.  

\begin{figure}[ctb]
\begin{center}
\includegraphics[width=3.4 in]{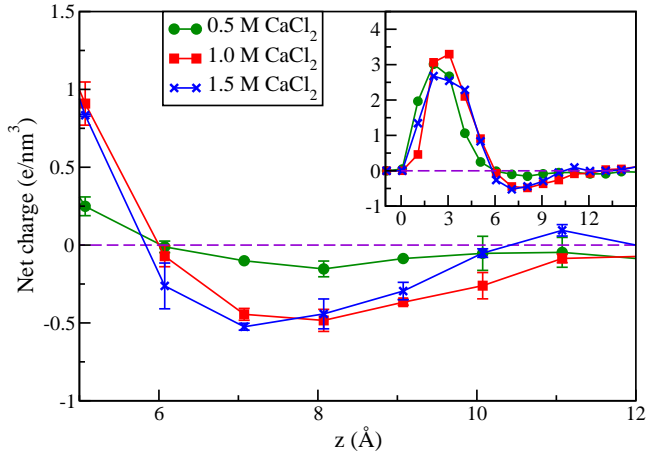}
\caption{(Color online:) Net charge density on the third layer
(see Fig.~\ref{fig:cartoon} for a definition of the layers) for
different concentrations of CaCl$_2$. The inset shows the profile of the
charge density.}\label{fig:charge_density}.
\end{center}
\end{figure}

The analysis of the time series Fig.~\ref{fig:time_series_T} and
Table~\ref{table:rhocharge} shows that the third layer has a net
negative charge, that is, there is an excess of Cl$^{-}$ charge
over divalent charge, which is the signature of CI. This is clear
from Fig.~\ref{fig:charge_density}, where the charge density as a
function of distance from the interface is shown. The results
exhibit no CI at 0.2 M (not shown) and marginal CI  for 0.5 M CaCl$_2$ , with CI
considerably amplified at larger concentrations. These results
show conclusive evidence for a Stern layer consisting of divalent
ions (first and second layers) followed by an ``inverted''
(negatively charged) diffuse layer. In all cases, charge densities
reach bulk values (within the accuracy of our simulations) at 12
\AA\ from the interface.

\begin{figure}[ctb]
\begin{center}
\includegraphics[width=3.4in]{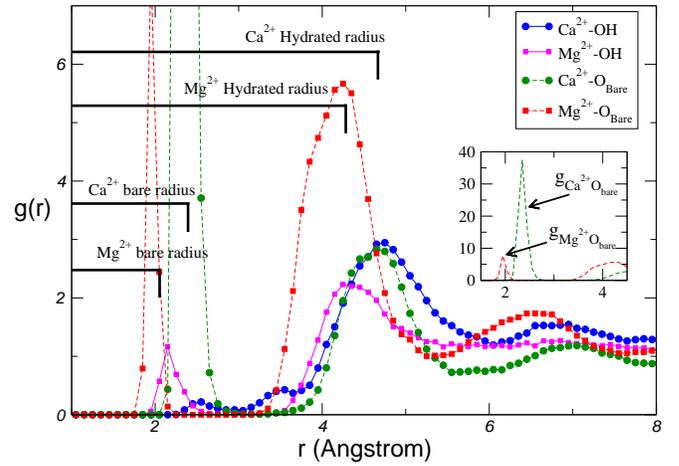}
\caption{(Color online:) Radials distribution function of
deprotonated silica oxygen atoms (O$_{bare}$) and protonated
oxygen atoms (OH) with Ca$^{2+}$. The inset
shows the rdf for (O$_{bare}$)-divalent ions at a larger vertical
scale.}\label{fig:rdf}.
\end{center}
\end{figure}

The structure of the ions near the silica interface is analyzed
using radial distribution functions (rdfs), shown in
Fig.~\ref{fig:rdf} for 1.0 M solutions. The rdf corresponding to
divalent ions with deprotonated silica oxygens (O$_{bare}$) show
sharp peaks at a distance roughly equivalent to the sum of the
crystallographic radius of oxygen and divalent ions providing
clear evidence for binding of partially dehydrated counterions to
oxygens. The inset of Fig.~\ref{fig:rdf} shows a much stronger
binding for Ca$^{2+}$ ions, sufficient to virtually neutralize the
interfacial charge, as clear from Table~\ref{table:rhocharge}. The
rdf also shows a maxima roughly corresponding to the sum of the radii of the hydrated divalent ion and an oxygen.  Although not bound, hydrated ions are strongly
correlated with silica oxygens, an effect that becomes more
dramatic with Mg$^{2+}$ ions. 

We further investigated the structure and orientation of water
near the interface and its relation to the observed ion
distributions. A detailed study of pure water near amorphous
silica interfaces will be reported elsewhere \cite{lorenz2006},
here we just briefly discuss the most salient aspects related to
ion distributions. It is found that the average lateral separation
of water molecules near charged silica are identical with pure
water results, but average distances between the water molecules and
oxygens on the silica interface are significantly altered
($4.0$~\AA\ as opposed to $3.2$~\AA\ for neutral silica), which we
interpret as being related to the Ca$^{2+}$ hydration sheaths pushing
away other water molecules. Our MD simulations do not show
evidence for any effect on the counterions attributable to a
pre-existing interfacial water ordering.

We briefly discuss the properties of the bulk solution as the
concentrations discussed in this paper cannot be considered as
dilute, and activity coefficients deviate considerably from unity.
It is found that the bulk rdf of Ca$^{2+}$-Cl$^{-}$ ions shows
strong correlations with a significant peak at a distance
corresponding to the sum of the crystallographic radius of both
ions (results not shown), thus providing clear evidence for
Bjerrum pairing. A quantitative analysis shows that the bulk
solutions contain $8(2)\%$, $17(5)\%$, and $25(7)\%$ of Bjerrum
pairs for concentrations of 0.5 M, 1.0 M and 1.5 M respectively.

\section{Discussion and Conclusions}

We now estimate the magnitude of the streaming current that would
be measured in the systems investigated. We consider a channel
with length $L$, width $w$ and height $h$ (see
Fig.~\ref{fig:cartoon}). The streaming current is
\begin{equation}\label{eq:I_str_def}
    I_{str}=w\int^{h/2}_{-h/2} dz \rho_c(z) v(z)
\end{equation}
where $\rho_c(z)$ is the net charge at position z and $v(z)$ is
the velocity of the fluid along the channel. This velocity is
given by Poiseuille flow $v(z)=\frac{\Delta P}{2\eta
L}(z^2-(\frac{h}{2})^2)$, where $\eta\sim 0.9$~cP is the shear
viscosity of water. Using the values of $\rho_c(z)$ from our
simulation, explicit values for the streaming current are
obtained. For wide silica channels ($h>>\lambda_D$), this
expression is further related to the $\zeta$-potential
\begin{equation}\label{eq:I_str_simpl}
    I_{str}=2 w\frac{\Delta P h}{2\eta L}\int^{\infty}_{0} dz z
    \rho_c(z)=-\Delta P \frac{w h \varepsilon_r}{4\pi \eta L}\zeta
\end{equation}
where $\rho_c(z)$ is the net charge distribution on an infinite
slab and $\varepsilon_r\sim 80$ is the dielectric constant of
water. The value of the $\zeta$-potential is defined at the plane
where the fluid velocity is zero, which we assume coincides with
the boundary between the Stern and the diffuse layer
\cite{Lyklema2006}. Using $h=140$~nm,$w=$50~$\mu$m and a length of the
channel $L=4.5$~mm as typical values \cite{Lemay2006}, we obtain
$I_{str}\approx 1$ pA/bar ($\zeta \approx 30$~mV) at
concentrations of $1$ and $1.5$~M (at $0.5$~M, $0<I_{str}<0.3
$~pA/bar).


This paper has presented all-atomic MD simulations of divalent
ionic solutions in silica channels. Our simulations show no CI at
$0.2$\ M so $0.2< c_{inv}<0.5$\ M. At concentrations larger than 0.5 M,
CI is substantial, both for CaCl$_2$ and MgCl$_2$. At
concentrations of 1.0 M, we calculate a streaming current of the
order of 1 pA/bar. These results compare extremely well with the
experiments in Ref.~\cite{Lemay2006}.

Our simulations provide a detailed understanding of the structure
of the ions near the interface and the underlying mechanism for
CI. It is shown that the Stern layer consists of both bound,
partially dehydrated (layer 1 or Inner Helmholtz plane), and
mobile hydrated counterions (layer 2 or Outer Helmholtz plane).
Counterions in layer 2 are strongly localized near the silica oxygens. In CaCl$_2$ solutions, the population
of the Inner Helmholtz Plane is significantly larger than the
Outer Plane.  Overall, our results support a picture where CI
is due to electrostatic correlations between counterions and
discrete interfacial groups (TC) or Bjerrum pairing
\cite{Travesset2006}. This is particularly clear in ions with
softer hydration sheaths, such as Ca$^{2+}$.

In summary, in this paper we provided a detailed account of the
mechanism of CI of divalent aqueous solutions near silica, in
excellent agreement with experimental results \cite{Lemay2006}.
By conducting simulations in which the water is explicitly modeled and the silica substrate is modeled in a realistic manner with discrete charges, we are able to capture the atomistic physics (i.e. structure of the ions and water near the charged species on the silica substrate) that plays such a large role in interfacial phenomena like that present in this problem.   It remains as a future challenge to develop simpler analytical models and further quantitative experimental tests.

{\bf Acknowledgments}

A.T. thanks S. Lemay  for inspiring discussions and R. Biswas for
insightful remarks on silica. C.L. acknowledges discussions with
E.\ B.\ Webb III, M.\ Stevens, G.\ S.\ Grest, M.\ Chandross, and M.\
Tsige. We both thank J.\ Anderson, J.\ Faraudo
and D.\ Vaknin for valuable discussions. A.T. wants to acknowledge the Aspen center for
physics for hospitality. This work is supported by NSF grant
DMR-0426597 and partially supported by DOE through the Ames lab
under contract no. W-7405-Eng-82.

\appendix
\section{}
\label{appendixa}
Table \ref{table:rhocharge_mgcl} shows the average populations of the Mg$^{2+}$ and Cl$^-$ ions in the three layers as defined in Fig.\  \ref{fig:cartoon}.  The values in Table \ref{table:rhocharge_mgcl} show that the first two layers have a net positive charge, and the third layer has a net negative charge.  This behavior is a signature of charge inversion and it is the same general behavior that was observed in the CaCl$_2$ systems.  The difference between the two systems is that in the MgCl$_2$ systems the silica charge is compensated by Mg$^{2+}$ in the second layer, whereas in the CaCl$_2$ systems the silica charge is compensated by Ca$^{2+}$ in the first layer.  

\begin{table}
\caption{Relative charge population of the ionic species of
Mg$^{2+}$ and Cl$^{-}$ in each layer.
Values in parentheses represent fluctuations in the last
digit(s).\label{table:rhocharge_mgcl}}
\begin{tabular}{|c|ll|ll|ll|}
\hline
System &1$^{st}$ layer& &2$^{nd}$ layer&&3$^{rd}$ layer&\\
  & Mg$^{2+}$ & Cl$^-$ & Mg$^{2+}$ & Cl$^-$ & Mg$^{2+}$ & Cl$^-$ \\
\hline
0.5 M & $0.26$ & $0.03(1)$ & $0.99(20)$ & $0.14(2)$ & $0.32(6)$ & $0.38(8)$ \\
1.0 M & $0.15$ & $0.08(1)$ & $1.54(39)$ & $0.43(11)$ & $0.65(8)$ & $0.84(13)$ \\
\hline
\end{tabular}
\end{table}

\begin{figure}[ctb]
\begin{center}
\includegraphics[width=3.0 in]{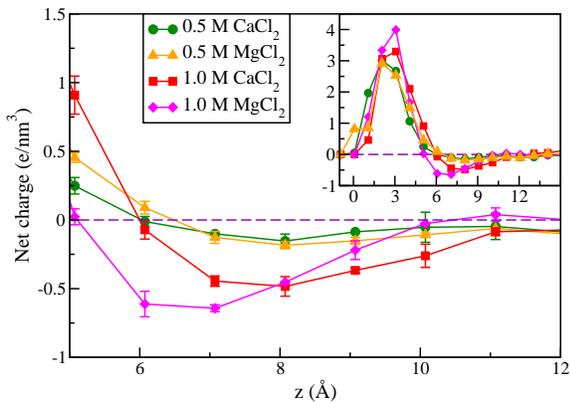}
\caption{(Color online:) Net charge density on the third layer
(see Fig.~\ref{fig:cartoon} for a definition of the layers) for
different concentrations of MgCl$_2$ and CaCl$_2$. The inset shows the profile of the
charge density.}\label{fig:charge_density_mg}.
\end{center}
\end{figure}

Charge inversion in the MgCl$_2$ systems is also obvious in Fig.~\ref{fig:charge_density_mg}, where the charge density as a function of distance from the interface is shown.  The data from the CaCl$_2$ systems are also plotted in Fig.~\ref{fig:charge_density_mg} so that they can be compared to the MgCl$_2$ systems.  In both cases, there is marginal CI of nearly identical magnitude at a concentration of 0.5 M.  Then at a concentration of 1.0 M, both systems show significantly larger CI.   These results for the MgCl$_2$ systems  show conclusive evidence for a Stern layer consisting of Mg$^{2+}$ ions (first and second layers) followed by an ``inverted'' (negatively charged) diffuse layer.

In MgCl$_2$ solutions, the number of divalent ions
within the Stern layer is similar as in CaCl$_2$ but its structure
is quite different, as there are very few bound counterions. We
raised the possibility that this may reflect that the Stern layer
may not be thermalized, and that longer simulations (at least $\sim$ 1
~$\mu$s) might reveal a structure more similar to the CaCl$_2$ case.
We point out, however, that in Ref.~\cite{Travesset2006} it was
concluded that binding constants of Mg$^{2+}$ to oxygen atoms are
generally small, a result attributed to the compactness of the
Mg$^{2+}$ hydration sheath. In any case, whether thermalized or not, our simulations show CI
and, as apparent from the rdf in Fig.~\ref{fig:rdf}, correlations
between Mg$^{2+}$ and charged interfacial groups are shown to play
a fundamental role.


\end{document}